\documentclass[preprint,proceedings]{rmaa}

\newcommand{\D}{\discretionary{}{}{}}

\SetYear{2010}
\SetConfTitle{IAU Latinamerican Regional Meeting}

\title{Fraction of Strong Barred Galaxies (SB) in the Nearby Universe, 0$\leq z\leq$0.066, as a function of redshift}

\author{
  J. A. Garc\'\i{}a-Barreto\altaffilmark{1}}

\altaffiltext{1}{Instituto de Astronom\'\i{}a,
  Universidad Nacional Aut\'o\-noma de M\'e\-xico, Apartado Postal
  70-264, M\'exico, D. F. 04510, M\'exico
  (tony\D{}@astroscu.\D{}unam.\D{}mx).}

\suppressfulladdresses

\listofauthors{J. A. Garc\'\i{}a-Barreto}
\indexauthor{Garc\'\i{}a-Barreto, J. A.}

\addkeyword{Galaxies: Barred}
\addkeyword{Galaxies: Redshift}
\addkeyword{Galaxies: Groups}
\addkeyword{Galaxies: Statistics}

\begin{document}
\maketitle 

\boldabstract{An analysis of 913 groups of galaxies and 56 clusters of galaxies from the literature
has been made in order to find the mean of the fraction of barred galaxies (SB/[S+SB]) and (SB/N) 
in the redshift interval 0$\leq z\leq$0.066.}

Our goal of our statistical study is to provide a reference fraction of strong barred (SB) 
galaxies {\bf in groups} in the optical in the nearby universe as a function of redshift 
($0\leq z\leq0.066$). How, when and at what rate did the barred galaxies form? 
This question is central to the field of (barred) galaxy formation and evolution. In order 
to investigate how disk and bar formation are related, it is not only important to determine 
the fraction of disk galaxies that have a bar (strong bar, classified as SB) but also to 
relate the bar and disk properties. Barazza, Jogee \&  Marinova (2008) have found a strong 
bar fraction SB/(S+SB) to be of the order of 48\% from a sample of 1144 disk galaxies from the SDSS.
Attempts to measure the bar fraction, SB/(S+SB), at high redshift ($0.2\leq z\leq$1) 
has been difficult. Sheth et al. (2008) using a detailed analysis of 2157 spiral galaxies in 
the interval from $0.2\leq z\leq0.84$ in the COSMOS 2 deg$^2$ survey concluded that the strong 
bar (SB only) fraction SB/(S+SB) decreses from about 27\% at $z\sim0.2$ to about 9\% at $z\sim0.835$. 

Our sample consists of 10,318 galaxies taken from 92 groups from the Nearby Catalog (Huchra \& Geller 
1982), from 176 groups from the CfA Catalog (Geller \& Huchra 1983), from 645 groups of the Tully 
Catalog (Tully 1987), and from 56 Abell clusters (Dressler 1980). Since each Abell cluster 
has more than 40 galaxies, they can be taken as defining an environment basis. This analysis then 
divided the groups in the three catalogs according to their number of galaxies. Thus the selection 
was done according to the number of galaxies in each group. For this, three environments were considered, 
dense groups having 10 or more galaxies (N$\geq10$), loose groups with $5\leq$N$\leq$9 galaxies and 
very poor groups with $2\leq N\leq4$ galaxies. Thus the statistics were done with 87 groups plus the 56 clusters 
(herein N$\geq10$), 103 groups with $5\leq$N$\leq$9 and 266 groups with N$\leq4$. 

Our results are: {\bf a)} SB/(S+SB) for the dense environments (N$\geq10$) is shown Fig. 1. It decreses from 43\% 
($0\leq z \leq 0.0099$), to 42\% ($0\leq z \leq 0.0129$), to 28\% ($0.001\leq z \leq 0.031$), to 
22\% ($0.011\leq z \leq 0.066$). {\bf b)} SB/(S+SB) for the loose groups ($5\leq N\leq9$) 
approximately stays constant from 33\% ($0\leq z \leq 0.0099$), to 29\% ($0\leq z \leq 0.0129$), to 33\% ($0.001\leq z \leq 0.031$). 
{\bf c)} SB/(S+SB) for very poor groups ($2\leq N\leq4$) stays constant at 50\% 
($0\leq z \leq 0.031$). Statistics of other fractions SB/N, (S+SB)/N, I/N, S0/N and 
E/N for the redshift considered are given in a forthcoming paper (Garc\'\i{}a-Barreto 2011).  

\begin{figure}[!t]
  \includegraphics[width=\columnwidth]{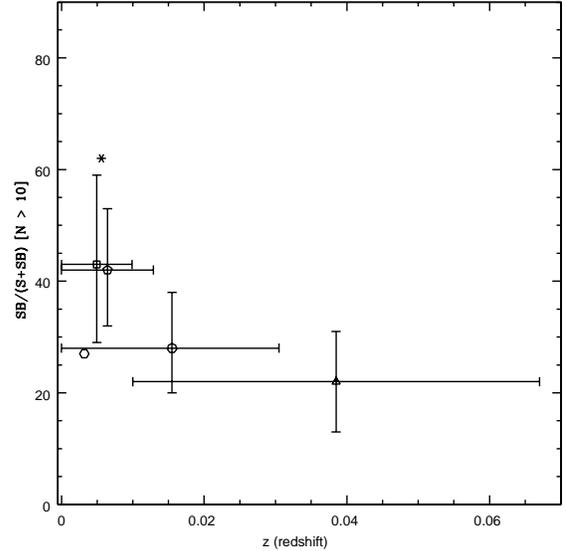}
  \caption{The fraction of strong barred galaxies, SB/(S+SB), in dense groups (N$\geq10$) 
decreases as a function of redshift $z$.}
  \label{fig:simple}
\end{figure}

\end{document}